\documentclass[12pt]{article}

\usepackage{amsfonts}

\usepackage{amssymb}
\usepackage{latexsym}
\usepackage{graphicx}
\usepackage[english]{babel}

\usepackage{amsfonts}
\usepackage{latexsym}
\usepackage{graphicx}
\usepackage[english]{babel}
\begin{document}
\input epsf

\def\p{\partial}
\def\h{{1\over 2}}
\def\be{\begin{equation}}
\def\bea{\begin{eqnarray}}
\def\ee{\end{equation}}
\def\eea{\end{eqnarray}}
\def\d{\partial}
\def\la{\lambda}
\def\eps{\epsilon}
\def\bb{\bigskip}
\def\mm{\medskip}
\newcommand{\dm}{\begin{displaymath}}
\newcommand{\edm}{\end{displaymath}}
\renewcommand{\b}{\tilde{B}}
\newcommand{\gm}{\Gamma}
\newcommand{\ac}[2]{\ensuremath{\{ #1, #2 \}}}
\renewcommand{\ell}{l}
\newcommand{\z}{\ell}
\newcommand{\newsection}[1]{\section{#1} \setcounter{equation}{0}}
\def\bb{$\bullet$}
\def\Qbar{{\bar Q}_1}
\def\QPbar{{\bar Q}_p}

\def\q{\quad}

\def\bn{B_\circ}

\let\a=\alpha \let\b=\beta \let\g=\gamma \let\d=\delta \let\e=\epsilon
\let\c=\chi \let\th=\theta  \let\k=\kappa
\let\l=\lambda \let\m=\mu \let\n=\nu \let\x=\xi \let\r=\rho
\let\s=\sigma \let\t=\tau
\let\vp=\varphi \let\vep=\varepsilon
\let\w=\omega      \let\G=\Gamma \let\D=\Delta \let\Th=\Theta
                     \let\P=\Pi \let\S=\Sigma

\def\h{{1\over 2}}
\def\t{\tilde}
\def\r{\rightarrow}
\def\nn{\nonumber\\}
\let\bm=\bibitem
\def\Kt{{\tilde K}}
\def\b{\bigskip}

\let\p=\partial

\begin{flushright}
\end{flushright}
\vspace{20mm}
\begin{center}
{\LARGE  What happens at  the horizon?\footnote{Essay awarded third prize in the  Gravity Research Foundation 2013 essay competition.}}
\\
\vspace{18mm}
{\bf  Samir D. Mathur }\\

\vspace{8mm}
Department of Physics,\\ The Ohio State University,\\ Columbus,
OH 43210, USA\\mathur.16@osu.edu\\
\vspace{4mm}
 March 31, 2013
\end{center}
\vspace{10mm}
\thispagestyle{empty}
\begin{abstract}

The Schwarzschild metric has an apparent singularity at the horizon $r=2M$. What really happens there? If physics at the horizon is `normal' laboratory physics, then we run into Hawking's information paradox. If we want nontrivial structure at the horizon, then we need a mechanism to generate this structure that evades the `no hair' conjectures of the past. Further, if we have such structure,  then what would the role of the traditional black hole metric which continues smoothly past the horizon? Recent work has provided an answer to these questions, and in the process revealed  a beautiful tie-up between gravity, string theory and thermodynamics.

\end{abstract}
\vskip 1.0 true in

\newpage
\setcounter{page}{1}

One of the most basic solutions to Einstein's equations is the Schwarzschild metric corresponding to a point source 
\be
ds^2=-(1-{2M\over r})dt^2+{dr^2\over 1-{2M\over r}}+r^2d\Omega^2
\label{one}
\ee
Near infinity this metric reproduces the expected weak field effects of a  mass $M$ placed at $r=0$. But moving inwards, we encounter a singularity at $r=2M$. What happens there? This question has led physicists through several twists and turns, and at the end, has led to a deep insight into the nature of quantum gravity. In this essay we recount this fascinating story, which has only recently reached its conclusion. 

\b

{\it First iteration:} \quad Particles falling in from infinity appear to slow down and freeze as they approach the horizon $r=2M$. Thus they never cross into the region $r<2M$ even if we wait till $t\r \infty$. This suggests the possibility that we may never need to talk about the region inside the horizon; physics should somehow be complete in the region $r>2M$. In the quantized theory of gravity, 't Hooft argued that there would be a `brick wall' at the horizon that scatters infalling quanta back to infinity \cite{thooft}. Susskind and his collaborators argued that quantum gravity effects would create an effective membrane at a `stretched horizon' just outside $r=2M$, where infalling quanta will be absorbed and reemitted \cite{susskind1}. If these views were correct, there would indeed be a complete description of black hole physics with no interior region $r<2M$. 

\b

{\it Second iteration:}\quad But the above picture soon runs into trouble. The singularity at $r=2M$ is just a coordinate singularity, and the metric can be continued smoothly across the horizon using Kruskal coordinates. Infalling particles appear to freeze at the horizon only because the time coordinate $t$ does not cover their full trajectory; when we switch to Kruskal coordinates then the particle trajectories continue through the horizon and reach $r=0$. 

The arguments for nontrivial effects at $r=2M$ in the quantum theory also turned out to be flawed.  In the Schwarzschild coordinate system we have $g_{tt}\r 0$ as we approach the horizon, and the corresponding time dilation leads to large quantum fluctuations for all fields. In particular the gravitational field has large fluctuations, and a naive analysis can suggest that something nontrivial is happening at the horizon. But since the horizon is a normal place in Kruskal coordinates, the effects of these violent quantum fluctuations should really all cancel out, leaving no `reflecting barrier' at the horizon. A closer analysis of the above mentioned quantum gravitational computations indeed indicates that they are  coordinate artifacts; see \cite{esko} for an example where the cancellations were demonstrated for a particular example.

Should we therefore accept that the horizon is a `normal place'? The problem with this conclusion is of course the information paradox pointed out by Hawking \cite{hawking}. If the vicinity of the horizon is a normal place where the local fields are in the vacuum, then we have a progressive creation of entangled Hawking pairs. As the black hole evaporates, the radiation at infinity gets progressively more entangled with the hole left behind. We then encounter  a sharp problem near the endpoint of evaporation where a tiny planck sized remnant must somehow be able to carry an arbitrarily large entanglement with the radiation at infinity. 

Most string theorists  had not worried too much about the information paradox, for a reason which turned out to be incorrect.  Since the number $N$ of emitted Hawking quanta is large, they assumed that small quantum gravity corrections of order $\epsilon\ll 1$ to the state of each created pair would be enough to make the overall state of the radiation unentangled from the remnant. But in \cite{cern} it was shown (using strong subaddditivity) that this belief was wrong; the reduction in entanglement entropy due to small corrections is bounded by
\be
{\delta S_{ent}\over S_{ent}}< 2\epsilon
\label{two}
\ee
Thus we can evade the information problem only by finding {\it order unity} corrections to evolution at the horizon. We get a `theorem': if the horizon is a `normal place' where lab physics holds to leading order, then we cannot evade the information paradox.

\b

\begin{figure}[htbp]
\begin{center}
\includegraphics[scale=.58]{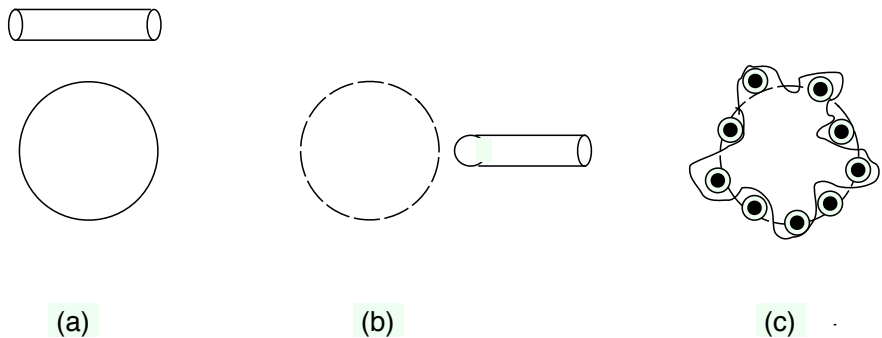}
\caption{{(a) Traditionally, it was assumed that in the black hole geometry the compact directions would appear as a trivial tensor product with the 3+1 metric. (b) In the actual microstates in string theory the compact directions pinch off to make KK monopoles/antimonopoles just outside the place where the horizon would have been. (c) The resulting solutions are `fuzzballs', which have no horizon or `interior'.}}
\label{feone}
\end{center}
\end{figure}

{\it Third iteration:} \quad The solution to the information problem is found by  actually constructing the states in string theory with mass $M$. It turns out that the geometry obtained is not the Schwarzschild one, but has the following structure (fig.\ref{feone}). For a hole in 3+1 dimensions, we have 6 compact directions, which we take to be small circles. One of these circles `pinches off' before reaching the horizon, so that spacetime ends there. There is an additional twist  that makes the metric near this pinch-off the metric of a KK monopole. The microstates of the hole are given by different configurations of KK monopoles and antimonopoles distributed around the rough location $r=2M$ where the horizon would have been, but in each case there is no horizon or `interior' of the horizon. These configurations are termed fuzzballs, and for simple holes it has been shown that all states of the hole are of this form \cite{fuzzball}.

Thus we have finally found `real' structure at the horizon, not a coordinate artifact. There are ergoregions between the KK monopoles, which radiate at exactly the rate expected from Hawking radiation \cite{radiation}. But there is no information paradox, since particle creation is not happening by the Hawking process where  one member of the created  pair fall through a horizon. 

This discussion suggests that the interior region $r<2M$ of the traditional black hole has no role at all. Susskind had postulated that there might be a 'complementary' description of the dynamics of the stretched horizon; in this description an infalling observer would see the traditional interior of the black hole \cite{susskind1}. In \cite{AMPS} the authors used the inequality (\ref{two}) to argue that such a complementarity would not be possible; the order unity corrections required at the horizon would create a `firewall' that cannot be consistent with a smooth continuation of the metric to $r<2M$. But as we will see now, there is a further twist to the story, so that the interior of the hole has a role even though no microstate actually has such an interior.

\b

\begin{figure}[htbp]
\begin{center}
\includegraphics[scale=.58]{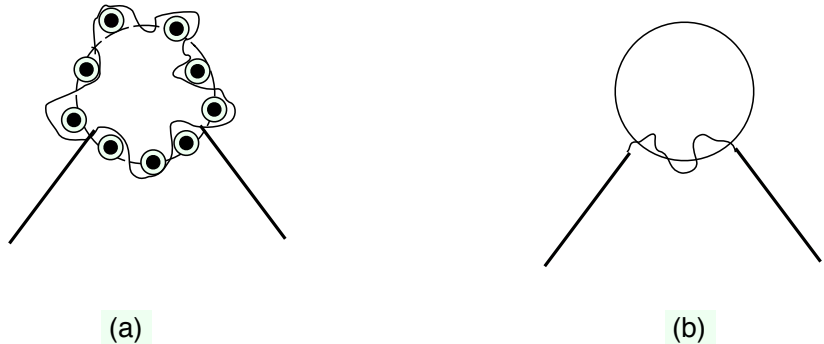}
\caption{{(a) Probing the fuzzball with operators at energy $E\gg kT$ causes collective excitations of the fuzzball surface. (b) The corresponding correlators are reproduced in a thermodynamic approximation by the traditinal black hole geometry, where we have no fuzzball structure but we use the geometry on both sides of the horizon.}}
\label{feetwo}
\end{center}
\end{figure}

{\it Fourth iteration:} \quad Consider probing the complicated surface of a generic fuzzball as shown in fig.\ref{feetwo}(a); this corresponds a 2-point function measured in a highly excited quantum gravity state. We will argue that, for suitable operators,  we can get a good approximation to such a correlator by using instead the traditional metric (\ref{one}) of the hole, as shown in fig.\ref{feetwo}(b). The latter geometry has none of the KK monopole excitations of the fuzzball surface but it does have the interior region $r<2M$. This is the sense in which we will find the role of the black hole interior.

\begin{figure}[h!]
\begin{center}
\includegraphics[scale=.58]{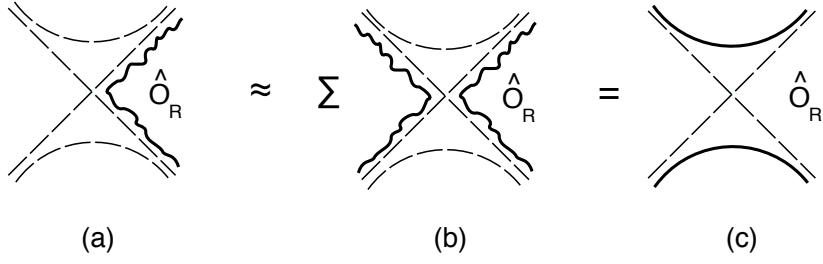}
\caption{{(a) Expectation value of $\hat O_R$ in one fuzzball; the geometry has only the region to the right of the fuzzball surface depicted by the wiggly line. (b) This  expectation value can be approximated by the ensemble average over fuzzballs. (c) The ensemble average is described by the traditional geometry with horizon.}}
\label{feethree}
\end{center}
\end{figure}

To motivate this proposal, recall the a scalar field $\phi$ on Minkowski space can be decomposed into fields in the right and left Rindler wedges.  Following arguments of Israel \cite{israel}, Maldacena \cite{maldacena} and van Raamsdonk \cite{raamsdonk}, we expect a similar decomposition where the state of an eternal black hole can be written as an entangled sum of gravitational states in the right and left quadrants:
\be
|g\rangle_{eternal}=C\sum_k e^{-{E_k\over 2T}}|g_k\rangle_L\otimes |g_k\rangle_R, ~~~C=\Big (\sum_i e^{-{E_i\over T}}\Big )^{-\h}
\label{qwfourt}
\ee
The states $|g_k\rangle_R$ live in the right wedge, and go to the vacuum at infinity. This is just the nature we observed for the fuzzball states, which asymptote to flat infinity and end just before reaching the horizon where a compact circle pinches off. Thus we conjecture that the states $|g_k\rangle$ are in fact the fuzzball states which describe microstates of the black hole \cite{plumberg}. We now make two observations (fig.\ref{feethree}):

\b

(i) The expectation value in the eternal black hole state of an operator in the right wedge is given by a thermal average over fuzzball states
\bea
{}_{eternal}\langle 0|\hat O_R|0\rangle_{eternal}&=&C^2\sum_{i,j}e^{-{E_i\over 2T}}e^{-{E_j\over 2T}}{}_L\langle g_i|g_j\rangle_L \, {}_R\langle g_i|\hat O_R|g_j\rangle_R\nn
&=&C^2\sum_i e^{-{E_i\over T}}{}_R\langle g_i|\hat O_R|g_i\rangle_R
\label{qwe1}
\eea

(ii) A given black hole is in {\it one} fuzzball state. But for a generic fuzzball state, and for suitable operators $\hat O_R$, we can approximate the expectation value by the ensemble average over all fuzzballs
\be
{}_R\langle g_k|\hat O_R|g_k\rangle_R\approx {1\over \sum_l e^{-{E_l\over T}}}\sum_i e^{-{E_i\over T}}{}_R\langle g_i|\hat O_R|g_i\rangle_R={}_{eternal}\langle 0|\hat O_R|0\rangle_{eternal}
\label{qwe2}
\ee
where in the second step we have used (\ref{qwe1}). But this is just the statement in fig.\ref{feetwo}, where the expectation value in one fuzzball is approximated by a black hole geometry which does have a region past the horizon.

\b

To summarize, we have arrived at the following picture. The microstates of the hole are the fuzzballs depicted in fig.\ref{feone}(c); these states have no horizon and no  `interior' region analogous to the region $r<2M$. The fuzzball radiates at the Hawking temperature $T$ from its surface just like any normal body, so there is no information  problem. Probing the fuzzball at energies $E\gg kT$ excites collective modes of the fuzzball which can be well approximated by an ensemble average over fuzzballs, and this average is reproduced by the traditional black hole geometry.

\end{document}